\documentclass[floatfix,showpacs,preprintnumbers,amsmath,amssymb,superscriptaddress,prb,twocolumn]{revtex4}

\usepackage{graphicx}
\usepackage{amssymb,amsmath}
\usepackage[dvips,usenames]{color}

\newcommand{\etalter}{{\it et al.}}
\newcommand{\eg}{{\it e.\ g.}}
\newcommand{\ie}{{\it i.\ e.}}
\begin{document}

\title{Fragility of the A-type AF and CE Phases of Manganites:\\
An Exotic Insulator-to-Metal Transition Induced by Quenched Disorder}
\author{G. Alvarez}
\affiliation{Computer Science \& Mathematics 
Division, Oak Ridge National Laboratory, \mbox{Oak Ridge, TN 37831}}
\author{H. Aliaga}
\affiliation{Condensed Matter Sciences Division, Oak Ridge National Laboratory, Oak Ridge, TN 32831}
\affiliation{Department of Physics and Astronomy, The University of Tennessee, Knoxville, TN 37996}

\author{C. \c{S}en}
\affiliation{National High Magnetic Field Laboratory and Department of Physics,
Florida State University, Tallahassee, FL 32310}

\author{E. Dagotto}
\affiliation{Condensed Matter Sciences Division, Oak Ridge National Laboratory, Oak Ridge, TN 32831}
\affiliation{Department of Physics and Astronomy, The University of Tennessee, Knoxville, TN 37996}

\begin{abstract} 
Using Monte Carlo simulations and the two $e_{\rm g}$-orbital model for manganites, the stability of
the CE and A-type antiferromagnetic insulating states is analyzed when quenched disorder
in the superexchange $J_{\rm AF}$ between the $t_{\rm 2g}$ localized spins 
and in the on-site energies is introduced. At
vanishing or small values of the electron-(Jahn-Teller)phonon coupling, the previously
hinted ``fragility'' of these insulating states is studied in detail, focusing on their
charge transport properties. This fragility is here found 
to induce a rapid transition from the insulator to a (poor) metallic state upon the introduction 
of disorder. A possible qualitative explanation
is presented based on the close proximity in energy of ferromagnetic metallic phases, and
also on percolative ideas valid at large disorder strength. The scenario is compared with previously 
discussed insulator-to-metal transitions in other contexts. 
It is argued that the effect unveiled here has unique properties that may define a new
class of giant effects in complex oxides. 
This particularly severe effect of disorder must be present in other materials as well, 
in cases involving phases
that arise as a compromise between very different tendencies, as it occurs with striped 
states in the cuprates.

\end{abstract}

\maketitle

\section{Introduction}\label{sec:intro}

The understanding of transition metal oxides is one of the most important open
challenges in Condensed Matter Physics. Exotic and complex phenomena emerge 
from the correlated nature of carriers in these materials, such as high critical 
temperature superconductivity and
colossal magnetoresistance (CMR).\cite{re:tokura00,re:dagotto05,re:salamon01,re:uehara99,re:dagotto01,re:mathur03,re:rao98,
re:deteresa97,re:lynn96,re:khomskii01,re:moreo99,re:ahn04,re:dagotto02,re:dagotto05b,
re:kumar06,re:verges02,re:salafranca06}  In recent years, much progress has been
made in this context after the realization that several competing states, with quite different
properties, are very close in energy in these compounds. As a consequence, small variations in
carrier density, temperature, pressure, magnetic fields, and other variables often lead to giant
responses and an intrinsic electronic softness.\cite{re:dagotto05} Moreover, the effect of quenched disorder appears to be important
to understand the CMR effect, according to recent theoretical and experimental
investigations briefly reviewed below. 
This combination of strong correlation and disorder effects, and the
simultaneous relevance of several degrees of freedom such as spin, charge, orbital,
and lattice, lead to a plethora of unusual properties, with tremendous potential
for functionalities. 

In this paper, using theoretical models and Monte Carlo simulations, the 
unexpected giant responses of some insulating states to the introduction of quenched disorder are discussed. Recent investigations 
by Aliaga \etalter\cite{re:aliaga03} already briefly described a curious ``fragility'' of the insulating
charge-ordered and orbital-ordered CE state of half-doped manganites. In those previous investigations the term
``CE glass'' was coined, and 
it was observed that the spin correlations at large distances characteristic of the CE state rapidly decrease 
in magnitude by 
increasing the strength of the quenched disorder in the links, namely by adding a random component to the
superexchange $J_{\rm AF}$ between the $t_{\rm 2g}$ spins degrees of freedom. 
In independent studies, Motome \etalter\cite{re:motome03c}
also observed that disorder suppresses 
charge order more strongly than ferromagnetic order using a one-band
model for Mn oxides (see also Sen \etalter\cite{re:sen04}). These conclusions appear to be
potentially important since experimental investigations by Akahoshi \etalter\cite{re:akahoshi03} and
Nakajima \etalter\cite{re:nakajima02} have observed that for a manganite material
in the clean limit (namely, with minimal quenched disorder) magnetoresistance effects
are absent. However, by introducing quenched disorder (via a random distribution of trivalent ions in the crystal) 
then a clear CMR effect is observed. Thus, experimental evidence suggests that quenched disorder is vital
to understand the manganites. This is in agreement with the early theoretical investigations of Burgy 
 \etalter~using simple ``toy'' models with two states in competition, supplemented by random resistor network
calculations.\cite{re:burgy01,re:burgy04} In those studies, quenched disorder was needed to obtain a 
CMR effect at low fields.

In the investigations presented in this manuscript, we have studied in considerable detail and using
larger lattices the fragility effect
first noticed in Refs.~\onlinecite{re:aliaga03,re:motome03c,re:sen04}. 
The focus of the investigations is on the half-doped regime of manganites,
where several previous efforts have established the phase diagrams with good accuracy. In particular, it is known that
a CE phase is clearly stabilized at intermediates values of $J_{\rm AF}$. Of much importance for the discussion
below is that this phase exists $independently$ of the value of the electron-(Jahn-Teller)phonon coupling $\lambda$.
Such a curious observation was understood independently by at least two groups 
(\eg, see Refs.~\onlinecite{re:hotta00,re:brink99}).
Hotta \etalter\cite{re:hotta00}  
provided an explanation where the well-known zigzag chains of the CE arrangement arise
from an optimization process between two competing tendencies:
the double exchange mechanism that favors ferromagnetism and the coupling between the
localized spins $J_{\rm AF}$ that favors antiferromagnetism. This competition leads to an exotic insulator, the
CE state, in a relatively narrow region of parameter space, interpolating between the more conventional ferromagnetic (FM) and antiferromagnetic (AF)
regimes. Ferromagnetic one-dimensional zigzag chains spontaneously emerge, with antiferromagnetic coupling between
these chains. Moreover, the insulating nature of this state arises from the unexpected $band$ $insulator$ properties of the
zigzag chains at $n$=0.5. This combination of frustrational and geometrical factors
leads to a stable insulating CE state. However, with hindsight clearly 
its nature can be anticipated to be quite fragile since its stability is induced by competition
of very different tendencies.

In our investigations, we have studied the CE state in two (2D) and three (3D) dimensions, as well as another insulating state, the
A-type AF state that is stable only in 3D. Small amounts of quenched disorder were introduced and we have analyzed
in detail the influence of this disorder over the transport properties of the resulting states, 
using canonical techniques.\cite{re:verges99} 
To our surprise, a
rapid transition from an insulator to a poor metal was observed. This effect appears robust: it is similar 
in both 2D and 3D,
and both for the CE and A-type insulators, to the extent that the electron-phonon coupling is small. In other words, it is
the ``topological'' CE insulator unveiled in previous investigations\cite{re:hotta00} that shows the strongest fragility toward
quenched disorder. This CE state is stabilized neither by a very strong nearest-neighbors Coulomb
repulsion nor by a very strong electron-phonon coupling, but by more subtle effects related to
frustration and geometry, thus causing its fragility. Also, these states appear to be highly susceptible to the 
introduction of magnetic fields, as also discussed in the text.

How exotic is the insulator-metal transition found here when analyzed in a more general framework?
Anderson localization transitions can be induced via the introduction of disorder in a metal.
In this case, quenched disorder transforms the metal into an insulator, which is the opposite 
of the transition we aim
to discuss in this paper. Recent investigations on the effects of correlated disorder have shown that
the metallic state can survive even in one dimensional systems,\cite{re:carpena02} 
but still the quenched disorder always favors the insulating tendencies, which does not happen in our case. 
In general, the appearance of a metal upon introducing disorder in an insulator is not a common phenomenon
if carried out at constant electronic density.
For instance, it is known that
insulator to metal transitions appear in conventional semiconductors. But they occur 
via doping and the concomitant creation of an impurity band. The transmission of charge
in disordered materials sometimes occurs via tunneling between large conducting regions.\cite{re:sheng80}
Doped polyacetylene has a
similar behavior, although the metallic phase is generated via a qualitatively different process.\cite{re:mele81}
A similar phenomenon occurs in other transition metal oxides, such as the high temperature
superconductors, where chemical doping introduces mobile carriers in the system,
rendering the compound metallic or even superconducting.\cite{re:dagotto05} In all these materials, carrier doping
away from the electronic density corresponding to the insulator is needed to achieve the metallic
state. This is not the case of the study reported in this paper that is carried out at a constant electronic
density $n$.  

Working at fixed $n$, an insulator to metal transition can occur also
if uniaxial stress increases the overlaps between otherwise localized electronic states. 
Our analysis is not of this variety, since the hopping
amplitudes (the model parameters which are the most susceptible to pressure) are left untouched in our investigations.

However, there are simple ways in which a disorder-induced insulator to metal transition can be envisioned
and it can occur by the
mere broadening of states by disorder. As the disorder strength increases, the gap will tend to close.
Examples of this behavior were discussed by Dobrosavljevi\'c and collaborators,\cite{re:tanaskovic03,re:vlad03} where 
dynamical mean-field theory calculations unveiled a metallic regime in between Mott insulator and
Anderson localized regions, at values of the disorder strength comparable to the Hubbard repulsion. 
However, we believe that this simple picture cannot be the full explanation of the numerical results reported
here since the metallic tendencies appear even for small values of the disorder strength. 
In the study of Ref.~\onlinecite{re:tanaskovic03} small amounts of disorder do not alter substantially the properties of the Mott
insulator, but the reciprocal occurs: small values of the Hubbard
U can  rapidly transform an insulator to a metal at intermediate values of the disorder strength.
Moreover, the subtle physics of zigzag chains is difficult to capture with mean-field approximations.
An explanation of the CE fragility as given in Ref.~\onlinecite{re:sen04} for the one-orbital model results of
Ref.~\onlinecite{re:motome03c} is also based on the simple broadening of peaks idea.
However, note that the CE state of half-doped manganites  at vanishing electron-phonon
coupling seems more fragile than other insulators, and we believe that a more sophisticated explanation
of its insulator to metal transition will be needed. 

Finally, to further motivate these investigations
it is important to remark that there is evidence suggesting that the CE state of the manganites
may actually be of the exotic ``topological''
variety discovered by Hotta \etalter\cite{re:hotta00}  In fact,
recent experimental efforts\cite{re:garcia01,re:daoud02} have challenged the common folklore that the
CE state contains ions with sharply-defined charge 3+ and 4+, 
since no evidence of strong charge order was observed. A
possible explanation of this puzzle
is provided in Fig.~10 of Ref.~\onlinecite{re:aliaga03} where the charge disproportionation between
the even and odd sites of the CE state is shown as a function of $\lambda$. It was there observed that the
regime of small $\lambda$ is the only one compatible with the new experiments. This is also in agreement
with the well known fact that a ferromagnetic metallic state must be very close in energy to the CE insulating
states, since experimentally the latter is easily destabilized by a magnetic field into the former. At large $\lambda$, 
the studies in Ref.~\onlinecite{re:aliaga03} showed that $all$ states (FM, CE, and AF) are charge ordered, in contradiction
with experiments. Thus, it is conjectured here that the fragility of the CE state may be of key relevance to
understanding the manganites. Recent experimental results are compatible with this conclusion, namely
quenched disorder affects the CE phase more aggressively than the FM state.\cite{re:mathieu04}

If we accept that the close proximity of states with different properties 
(such as a metal or superconductor and an insulator) is the basic cause of the effect, then
it follows that 
the giant effect of
quenched disorder should be present in many other systems as well. For instance, the famous 
stripes\cite{re:tranquada95,re:kivelson03} in
cuprates have recently been found to appear at intermediate narrow coupling regimes of
phenomenological models for the superconductor-antiferromagnetism
competition,\cite{re:alvarez05c,re:mayr06} namely they are also a compromise between very different states. As a consequence, we here
conjecture that Cu-oxide stripes should also be highly susceptible to the explicit introduction of disorder.
This conclusion can be extended to any intermediate narrow phase of transition metal oxides that originates from
competing tendencies.

The organization of this manuscript is the following. In Section II, the model is described. Since this
model has been extensively used in previous investigations, the discussion is brief. Section III contains
the main results, namely the influence of link disorder on the transport properties of the system. The
analysis is carried out for two insulating states (the CE and A-type AF phases) and in both two and three
dimensions. Evidence for the metalization of the system upon the introduction of link quenched
disorder is presented. Section IV describes the results for transport now including magnetic fields.
Section V deals with on-site disorder, Section VI provides evidence that our results are not dramatically
altered by increasing lattice sizes, and finally conclusions are provided in Section VII.

\section{Model and Details of the Simulation}

The standard double-exchange
two-orbital model for manganites was used in this study. Since
the model has been widely described in previous 
investigations,\cite{re:dagotto01} this section is brief.
 The two orbitals arise from the $e_{\rm g}$ states that are active at the Mn ions in Mn-oxides,
as extensively discussed before.\cite{re:dagotto01,re:dagotto02,re:dagotto05b} 
The electronic density will be $n$=0.5 in all the results reported here. This
corresponds to one electron per site of a square or cubic lattice, and it is
the density where the CE state of manganites is experimentally known to be stable.

In more detail, the Hamiltonian for the two-orbital model is
\begin{eqnarray}
H&=&\sum_{\gamma,\gamma',i,\alpha}t^\alpha_{\gamma\gamma'}
{\mathcal S}(\theta_i,\phi_i,\theta_{i+\alpha},\phi_{i+\alpha})
c^\dagger_{i,\gamma}c_{i+\alpha,\gamma'} \nonumber \\ &+& 
\lambda\sum_{i}(Q_{1i} \rho_i +  Q_{2i} \tau_{xi} + Q_{3i} \tau_{zi}) +
\sum_{i}\sum_{\alpha=1}^{\alpha=3} D_\alpha Q_{\alpha i}^2\nonumber \\&+&
 \sum_{i,\alpha} J_{\rm AF}(i,\alpha)\vec{S}_i\cdot
\vec{S}_{i+\alpha}+\vec{B}\cdot \sum_i \vec{S}_i,
\label{eq:hamtwobands}
\end{eqnarray}
where the factor that renormalizes the hopping in the $J_{\rm H}\rightarrow\infty$ limit is
\begin{eqnarray}
{\mathcal S}(\theta_i,\phi_i,\theta_{j},\phi_{j})&=&\cos(\frac{\theta_{i}}{2})\cos(\frac{\theta_{j}}{2})
\nonumber \\&+&\sin(\frac{\theta_{i}}{2})\sin(\frac{\theta_{j}}{2})e^{-i (\phi_{i}-\phi_{j})},
\end{eqnarray}
and the angles are used to parametrize the $t_{\rm 2g}$
classical spins at every site.
The parameters $t^\alpha_{\gamma\gamma'}$ are the 
hopping amplitudes 
between the orbitals $\gamma$ and $\gamma'$ in the direction $\alpha$. In 
two dimensions, these are $t^{x}_{aa}=-\sqrt{3}t^{x}_{ab}=
-\sqrt{3}t^{x}_{ba}=3t^{x}_{bb}=1$,
and $t^{y}_{aa}=\sqrt{3}t^{y}_{ab}=\sqrt{3}t^{y}_{ba}
=3t^{y}_{bb}=1$ and in three dimensions one needs also consider $t^{z}_{aa}=t^z_{ab}=t^z_{ba}=0$ and
$t^z_{bb}=4/3$. 

Although the phononic coupling is barely used in the effort reported
here, for completeness a description if provided.
$Q_{1i}$, $Q_{2i}$ and $Q_{3i}$ are normal 
modes of vibration that can be expressed in terms of the oxygen coordinate $u_{i,\alpha}$ as:
\begin{eqnarray}
Q_{1i}&=&\frac {1}{\sqrt{3}} [(u_{i,z}-u_{i-z,z}) + (u_{i,x}-u_{i-x,x}) \nonumber \\ &+&  
(u_{i,y}- u_{i-y,y})], \nonumber \\
Q_{2i}&=&\frac {1}{\sqrt{2}} (u_{i,x}-u_{i-x,x}), \nonumber \\
Q_{3i}&=&\frac {2}{\sqrt{6}} (u_{i,z}-u_{i-z,z}) 
 -\frac{1}{\sqrt{6}} (u_{i,x}-u_{i-x,x}) \nonumber \\&-&\frac{1}{\sqrt{6}} (u_{i,y}- u_{i-y,y}). \nonumber
\end{eqnarray}
Also, $\tau_{xi}$=$c^{\dagger}_{ia}c_{ib}$+$c^{\dagger}_{ib}c_{ia}$, 
$\tau_{zi}$=$c^{\dagger}_{ia}c_{ia}$-$c^{\dagger}_{ib}c_{ib}$, and 
$\rho_{i}$=$c^{\dagger}_{ia}c_{ia}$+$c^{\dagger}_{ib}c_{ib}$.
The constant $\lambda$ is the electron-phonon coupling
related to the Jahn-Teller distortion 
of the MnO$_6$
octahedron.\cite{re:dagotto01}
Regarding the phononic stiffness, and in
units of $t^{x}_{aa}$=$1$, the $D_{\alpha}$ parameters are $D_1$=$1$ and 
$D_2$=$D_3$=$0.5$,
as discussed in previous literature.\cite{re:aliaga03} 
The rest of the notation is standard, including a superexchange term with coupling $J_{\rm AF}$, which can be made
spatially inhomogeneous due to the introduction of quenched disorder in the form of a bimodal distribution of
amplitude $\Delta$ (that will always be expressed in units of $t^x_{aa}$), namely at each link the local coupling
$J_{\rm AF}(i,\alpha)$ is either $J_{\rm AF}+ \Delta$ or $J_{\rm AF}- \Delta$
with equal chance. This type of disorder is natural if the effect of chemical
doping in manganites is considered, where ions of a given size are replaced
by ions of a different size creating local stress in the crystal.
Also, in some of the investigations reported here, 
a Zeeman term with field strength $B$ was added to investigate
the influence of magnetic fields.

The two orbital model was here studied  mainly using the canonical Monte Carlo method for spin-fermion
systems, widely described in previous literature.\cite{re:dagotto01} The conductances were
calculated using a method proposed by Verges\cite{re:verges99} and the unit of conductance throughout this paper
is $e^2/h$. The lattices studied here
are of intermediate size (with the exception of those in Section VI using the TPEM), but note that
the computational effort was substantial (see below) in view of the many parameters in the problem, plus the
need of averaging over several configurations of disorder. Future work exclusively focused on
a small region of parameter space could address even larger systems.

It is also important to remark that the computational work described
in this manuscript has several unique characteristics, including the
vast resources that it has required to accurately simulate the
manganite models. To give the reader a perspective of the effort,
note that most of the simulations have been carried out on
the NCCS XT3 supercomputer (with 2.4-GHz AMD Opteron processors and 2 GB of
memory), using 100 nodes in parallel for small systems
and up to 1600 nodes per run on large lattices.
The entire effort documented in this paper would have
required several years using a cluster of commodity PCs.
Parallelization has been implemented with the MPI at three levels.
First, the integration of the electronic sector, which is carried out with the
TPEM, has been parallelized. The next level is related to the presence of quenched
disorder in the system that requires the use of many different disorder
configurations. Finally, the simulations have to sweep over a large
parameter space ($\lambda$, $J_{\rm AF}$ and temperature) allowing for a
third level of parallelization.

\section{Effect of Link Disorder on the Resistance}

\begin{figure}
\centering{
\includegraphics[width=5cm,clip]{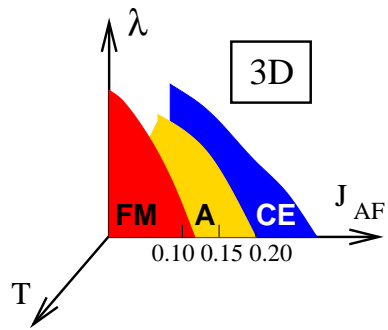}
}
\caption{\label{fig:phase3dsketch} Schematic representation of the phase diagram corresponding to the model
Eq.~(\ref{eq:hamtwobands}) in three dimensions,\cite{re:aliaga05} and electronic density
$n$=0.5. The influence of quenched disorder on the conductance 
will be mainly studied in this case in the $\lambda=0$ plane (\ie without electron-phonon
coupling), focusing on the A-type and CE phases.}
\end{figure}

In this section, the effect of quenched disorder (in the $J_{\rm AF}$
link variables)
on the insulating phases 
that appear in models for manganites will be investigated. Our main goal is to analyze if
these insulators are robust, namely if they preserve their properties
with increasing disorder, or whether they are fragile under the
same ``perturbation''. Surprisingly, the latter is found to be the case:
large changes in the properties of the insulating state occur even for fairly modest
disorder strengths. This is in good agreement with several experiments 
in the manganite context, as explained before. 
Our results provide evidence that some states, such as the CE,
are fragile, namely easily disturbed by perturbations.
The study in this section is organized following the sketch given in
Fig.~\ref{fig:phase3dsketch} for a 3D case. 
First, the A-type antiferromagnetic state will be studied, 
followed by the CE phase. 
These are the closest states to the important FM metallic state.
At the end of the
section, the 2D CE state is also analyzed.

\subsection{A-type antiferromagnet in 3D}

The A-type AF phase consists of ferromagnetically ordered planes, that are 
antiferromagnetically stacked along one axis. At $\lambda$=0 and in 3D, this
state is stable at intermediate values of $J_{\rm AF}$, as shown in 
Ref.~\onlinecite{re:aliaga05}.
It is interesting to observe that this
phase is obtained in our study using unbiased Monte Carlo simulations\cite{re:dagotto01,re:kajimoto02} 
in the clean limit ($\Delta$=0)
starting, for instance, from a fully paramagnetic (random) classical spin configuration. 
There are no problems identified here of convergences 
to higher energy states.
However, note that due to its intrinsic anisotropy, the conductance of the A-type AF phase is 
highly directional, \ie, the system is metallic in the FM
plane but insulating when the conductance is measured along the direction 
where the system is AF. In Monte Carlo simulations, while the A-type character of the
state is always obtained in the appropriate range of parameters, the orientation (\ie the
antiferromagnetic direction) is random.
This is clear since the three orientations are energetically degenerate.
Since our goal is to study the A-type AF insulator, as it appears in experiments where it
has an insulating character,
then we measure the conductance along the insulating direction of the A-type phase.

\begin{figure}
\centering{
\includegraphics[width=8cm,clip]{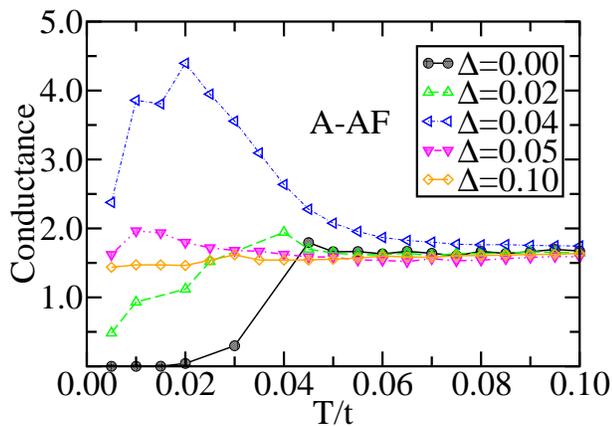}
}
\caption{\label{fig:atyperes1} Results illustrating the appearance of a metal upon adding quenched disorder
to an insulator. Shown is the
numerically calculated conductance vs. temperature $T$ for the A-type AF phase in three dimensions, with the antiferromagnetic
spin arrangement along the $z-$axis, with and without disorder. 
The calculation was done on a $4^3$ lattice, working 
at $J_{\rm AF}=0.15$ and $\lambda=0$. For each temperature and $\Delta$, 5,000 thermalization and 15,000 measurement Monte
Carlo steps through the whole lattice were carried out. The actual measurements of observables were separated by 10 steps.
The starting configuration corresponded to an A-type AF in the $z$-direction. 
The link disorder is bimodal, $J_{\rm AF}\pm\Delta$, and the average shown is over
5 quenched disorder configurations. For the error related to this average see 
Fig.~\ref{fig:atyperesd}a.}
\end{figure}

In Fig.~\ref{fig:atyperes1} the conductance vs. temperature, $T$, is plotted for various 
disorder strengths, $\Delta$. The study is carried out at $J_{\rm AF}=0.15$,
which we found corresponds to an A-type antiferromagnetic ground state for $\Delta=0$. 
Shown are averages over 5 configurations of quenched disorder. At $\Delta$=0, the conductance
vanishes at low temperatures, result compatible with the expected insulating properties
of the A-type AF phase (the spin-spin correlations and spin structure factor indeed confirm
the A-type spin character of this state).

The most interesting results of Fig.~\ref{fig:atyperes1} are obtained at nonzero $\Delta$.
Results at $\Delta$=0.02 indicate an enhancement of metallicity (higher conductance)
than in the clean $\Delta$=0 limit, which is surprising. 
The system has a $finite$ conductance in this regime. At $\Delta$=0.04,
(a disorder strength that can be considered intermediate with respect to the 
central value of the disorder distribution, $J_{\rm AF}=0.15$), 
the effect reported in this paper
is the most dramatic. At this value of $\Delta$, the low temperature conductance reaches
a maximum.  Upon further increasing $\Delta$, the transport properties now become that
of a poor metal, namely the actual conductance value is low (at least compared with
its maximum possible value), and the result is nearly temperature independent. It is the
surprising regime of small/intermediate $\Delta$ that it is the most interesting due to its dramatic
change in conductance, but the large $\Delta$ region is also puzzling since at very large
disorder one would expect to find an insulator.

\begin{figure}
\centering{
\includegraphics[width=8cm,clip]{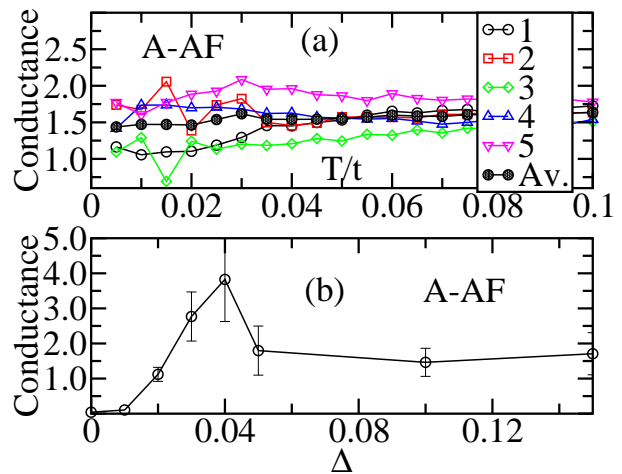}
}
\caption{\label{fig:atyperesd} 
(a) Dependence of the conductance on the particular
quenched-disorder configuration used (labeled as 1-5, with ``Av.'' denoting the average). 
Although the results near the metal-insulator
transition vary substantially from configuration to configuration, the qualitative
behavior is similar in all of them.  Shown is the conductance vs. $T$ for the A-type AF phase (AF along the $z$-axis)  for
different disorder configurations, using $\Delta=0.1$. 
The calculation was done on a $4^3$ lattice at $J_{\rm AF}=0.15$ and $\lambda=0$, with the same
number of Monte Carlo steps discussed in Fig.\ref{fig:atyperes1}.
The disorder is bimodal, $J_{\rm AF}\pm\Delta$, and the average shown is 
over the 5 disorder configurations.
(b) Dependence of the conductance on the disorder strength $\Delta$, at low temperatures, showing the
transition from an insulator to a metal upon disordering the system. Shown is the
conductance vs. $\Delta$ of the A-type AF phase (AF along the $z-$axis) at $T=0.02t$.
The lattice, couplings, Monte Carlo steps, initial configuration, and type of disorder used are as in
(a). The error bars mainly arise from the average over several
quenched disorder configurations.
}
\end{figure}

In studies incorporating quenched disorder, the averages over disorder configurations are
important. 
Indeed, disorder introduces another source of error (apart from Monte Carlo error) and the
size of this error needs to be determined.
To that effect the dependence of 
the conductance on each of the five different disorder
configurations used in the previous figure is shown in Fig.~\ref{fig:atyperesd}a for 
$\Delta=0.1$. 
 We found an analogous 
behavior for other values of $\Delta$ although this error tends
to be larger for smaller $\Delta$s. Moreover,
a similar dependence of the conductance on the disorder configuration 
was found for the CE phase.

The low temperature ($T$=$0.02$) dependence of the conductance on the disorder strength, $\Delta$, is
shown in Fig.~\ref{fig:atyperesd}b. It is remarkable that the connection between the insulating
state at $\Delta$=0 and the large $\Delta$ limit is not smooth, but a peak is observed
at intermediate disorder strength.

\begin{figure}
\centering{
\includegraphics[width=7cm,clip]{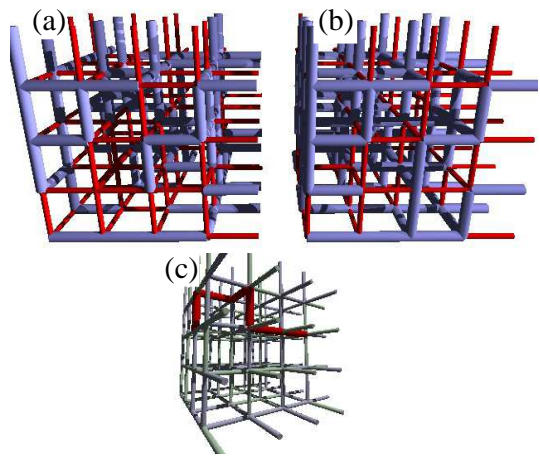}
}
\caption{\label{fig:cubes}
Figures illustrating the expected dominance of quenched disorder 
over the spin configurations, at large $\Delta$.
(a) Sign of $\vec{S}_i\cdot\vec{S}_j$ for nearest-neighbor sites $i,j$, 
indicated as red/thin (negative sign; antiferromagnetic) 
and blue/thick (positive sign; ferromagnetic) connections. The result corresponds to
one particular quenched disorder and classical spin configurations, obtained
after the Monte Carlo thermalization. 
(b) Value of the link coupling 
$J_{\rm AF}(i,j)$, indicated as colors red/thin ($J_{\rm AF}+\Delta$) and blue/thick ($J_{\rm AF}-\Delta$). In both
left and right panels, the couplings were $J_{\rm AF}$=$0.15$ and 
$\Delta$=$0.1$.
The correlation between the results in both panels is apparent, indicating that for this large
value of the disorder strength the classical spins follow the disorder in the spin-spin correlations.
(a) An example of a FM percolative path found in the study of the top panels. There are
many of them (but only one shown).
}
\end{figure}

A possible explanation of the results obtained in this study is the following. Let us start at very large
$\Delta$, namely in the regime where disorder dominates. It may appear unnatural that in such a limit
a metal (even if poor) is obtained, since disorder is associated with localization and related 
effects. However, in the large $\Delta$ regime each of the lattice links 
(defined by two nearest
neighbor sites) has an associated
pair of classical spins which are either mainly ferromagnetic or antiferromagnetically aligned, depending
on the value of the coupling $J_{\rm AF}$ at that particular link. If the link coupling is
$J_{\rm AF}$+$\Delta$, then this link is AF, while for $J_{\rm AF}$-$\Delta$ it tends to be
ferromagnetic due to the double exchange mechanism. The fact that at large $\Delta$ the orientation of the
classical spins at each link follows the disorder is clear from Fig.\ref{fig:cubes} (top panels)
where both the spin correlations and the $J_{\rm AF}$ couplings
are shown. Although a few links have different color (or thickness) on the left and right top panels,
the majority coincide, namely the spins follow the couplings. 
The important point in this regime is that the ferromagnetic links can percolate
since there are (on average) 50\% of them, and in three dimensions the critical concentration for percolation is
only 25\%. A percolative path is shown in the lower panel of Fig.\ref{fig:cubes}.
However, note that 
this explanation is only rough, since for Heisenberg classical
variables (\ie defined by two angles) as in our simulations 
the correlation between two links does not necessarily take two values (FM/AF) as in the case of Ising spins.
Nevertheless, we have observed that even within the many percolative channels, 
the spins tend to twist their orientations to satisfy better the competing FM-AF tendencies
reducing the conductance substantially. 
The situation would be different in systems with sufficient anisotropy to be described by Ising variables.
The important fact is that even if the FM links are percolated, the AF links provide a frustrating tendency to
them, \ie they are not independent of each other. Thus, the resulting conductance is finite but poor.
Nevertheless, the essence of the metallic behavior found at large $\Delta$ is related to
the percolative picture described in this paragraph.

For low and intermediate disorder strengths, the conductance is larger than at large $\Delta$.
A different, more complex explanation of the results holds in this regime, and
this will be discussed later in this paper, after confirming that the large intermediate conductance
occurs in other systems as well.

\subsection{CE phase in 3D}

\begin{figure}
\centering{
\includegraphics[width=8cm,clip]{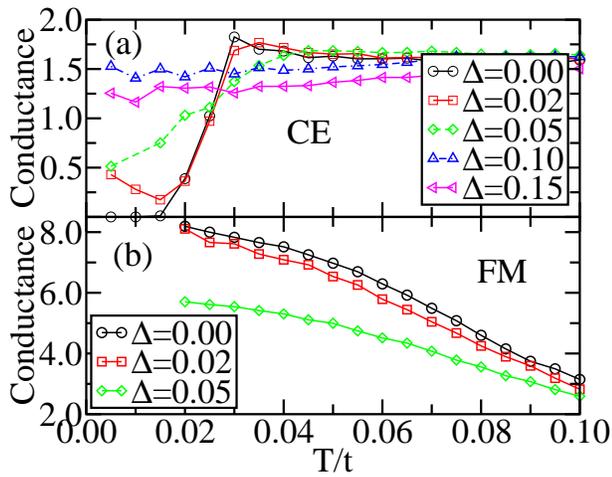}
}
\caption{\label{fig:cefmdisorder3d} 
(a) Results illustrating the metal to insulator transition induced by disorder, focusing on 
the regime of the CE phase in a three-dimensional lattice.
Shown are the conductance vs. $T$ of the CE-phase,  with and without quenched disorder of strength $\Delta$. 
The calculation was done on a $4^3$ lattice, at $J_{\rm AF}=0.20$, $\lambda=0$, and using
5,000 for thermalization and 15,000 for measurement Monte Carlo steps (the latter actually evaluating
observables at every 10 steps to avoid self-correlations). 
The initial starting configuration had a CE phase with zigzag chains in the xy-plane. 
The disorder is bimodal, $J_{\rm AF}\pm\Delta$, and shown are averages over
5 disorder configurations.
(b) Conductance vs. $T$ for the FM phase,  with and without disorder. In this case, where a metal exists at
$\Delta$=0, the disorder simply decreases the conductance, as intuitively expected.
The lattice, Monte Carlo steps, number of disorder configurations,
and type of link disorder are as in (a). The
superexchange coupling is $J_{\rm AF}$=$0.10$.
 }
\end{figure}

An investigation similar to that reported in the previous section for the A-type AF phase was
also carried out for the CE phase. The CE state is stabilized in the clean limit by simply further increasing
$J_{\rm AF}$ with respect to the value used for the A-type AF states, 
as reported in previous investigations (see sketch Fig.\ref{fig:phase3dsketch}).
In Fig.~\ref{fig:cefmdisorder3d}(a) the conductance vs.~$T$ is presented for $J_{\rm AF}$=$0.2$ (corresponding to a
CE-phase ground state at $\Delta$=0), and for various strengths of the disorder configuration. At $\Delta$=0, the
expected result is found, namely for temperatures below a critical one the conductance vanishes. This clearly
signals an insulator, which has CE characteristics based on the values of the spin and charge 
correlations.\cite{re:dagotto02}
For small values of $\Delta$, the conductance increases from zero at low temperatures, and at $\Delta$=0.10 it
reaches a plateau, which is nearly temperature independent. Note that contrary to the results for the A-type AF
phase, the conductance grows monotonically until the large $\Delta$ (percolated) regime is reached.
It will be argued later that this is caused by the fact that the FM metallic phase is no longer a neighbor
of the CE state in the 3D phase diagram. 
For completeness,
the case of a FM ground state (stabilized at smaller values of $J_{\rm AF}$)
is also shown in Fig.~\ref{fig:cefmdisorder3d}(b). At all temperatures, the conductance decreases
with increasing disorder strength. For large enough $\Delta$ (not shown), a common limit is reached both in
the cases of the (a) and (b) panels in the figure. The metallic FM state is damaged by disorder and its conductance
is reduced, as intuition
suggests, while the insulating states (A or CE type) actually 
become metallic after the addition of disorder.

\subsection{CE phase in 2D}\label{subsec:ce2d}

\begin{figure}
\centering{
\includegraphics[width=5cm,clip]{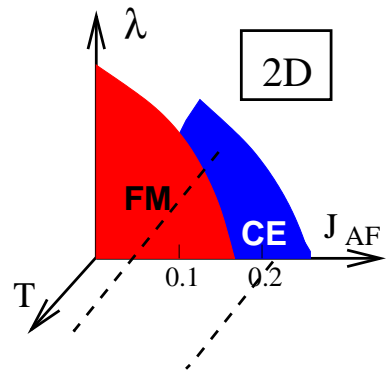}
}
\caption{\label{fig:phase2dsketch} Schematic phase diagram for model
Eq.~(\ref{eq:hamtwobands}) in two dimensions, from Ref.\onlinecite{re:aliaga03}. The effect of quenched
disorder is studied in this section
in the cases: (i) $\lambda=0$, $J_{\rm AF}$=$0.17$, corresponding to a CE phase without active 
phonons, and (ii) $\lambda=1$, $J_{\rm AF}$=$0.145$, corresponding to a CE phase stabilized including phonons.
These two cases are indicated by dashed lines.}
\end{figure}

To verify the universality of the rapid transformation of the CE phase into a poor metal upon
introduction of disorder found in 3D, similar calculations have been carried out in 2D.
In this case, a plethora of previous investigations \cite{re:dagotto01} have shown that the CE phase is stabilized
at $n$=0.5 by increasing the coupling $J_{\rm AF}$. A typical phase diagram is sketched
in Fig.~\ref{fig:phase2dsketch}. Note that the CE phase is stabilized even at $\lambda$=0,
mainly for topological reasons related with the kinetic energy of carriers in zigzag chains.\cite{re:hotta00}
However,
in the analysis below, we will also study the CE phase in
the regime where the electron-phonon coupling $\lambda$ is important, to search for 
qualitative differences.

\begin{figure}
\centering{
\includegraphics[width=8cm,clip]{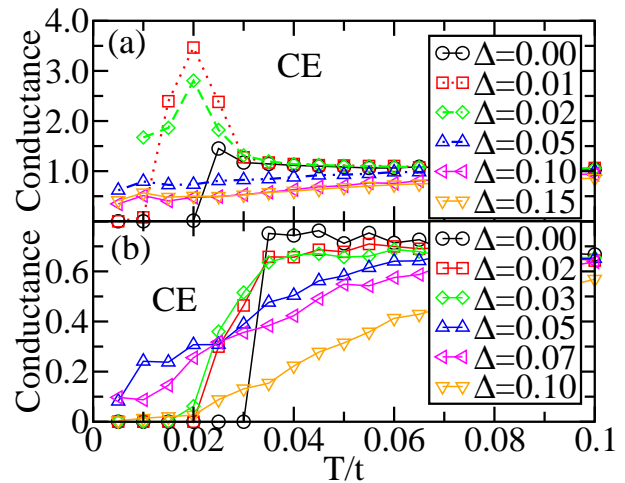}
}
\caption{\label{fig:cejaf} Insulator to metal transition induced by quenched disorder in two dimensions.
(a) Shown is the conductance vs. $T$ of the
two-dimensional CE phase,  for different disorder strengths $\Delta$. 
The calculation was done on an $8\times8$ lattice, at $J_{\rm AF}$=$0.17$, and $\lambda=0$. 
The number of Monte Carlo steps was 5,000 for thermalization and 15,000 for measurements, calculating
observables every 10 steps. The initial configuration was the perfect CE phase. 
The disorder is bimodal, $J_{\rm AF}\pm\Delta$, and the averages shown are
over 5 quenched disorder configurations. (b) Same as (a) but for $J_{\rm AF}=0.145$ and  $\lambda=1.0$.
}
\end{figure}

\begin{figure}
\centering{
\includegraphics[width=8cm,clip]{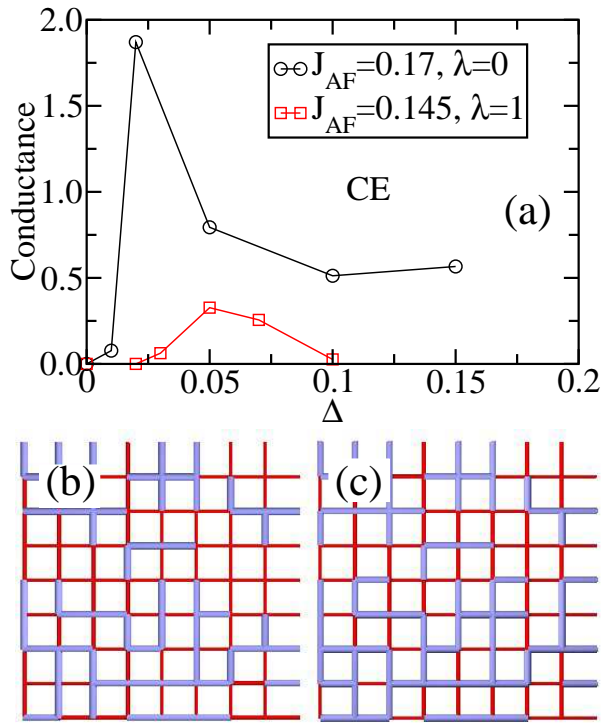}
}
\caption{\label{fig:l8x8cvsd} (a) Dependence of the conductance on disorder strength $\Delta$, 
at low temperature $T$=0.01, using results taken from Fig.~\ref{fig:cejaf}(a).
The most dramatic evidence of 
fragility in the 2D insulating CE state is obtained in the absence of electron-phonon coupling.
(b) Sign of $\vec{S}_i\cdot\vec{S}_j$ for nearest-neighbor sites $i,j$, indicated as red/thin
(negative; AF) and blue/thick (positive; FM) connections. (c)
Value of $J_{\rm AF}(i,j)$ indicated as two colors red/thin ($J_{\rm AF}+\Delta$) 
and blue/thick ($J_{\rm AF}-\Delta$). In both
cases, $J_{\rm AF}=0.17$, $\Delta$=$0.1$, and the results were obtained for fixed classical 
spin and quenched disorder configurations.
The correlation between the quantities in the (b) and (c) panels is apparent, 
indicating that for this large
value of $\Delta$ the system follows the disorder in the spin-spin correlations.}
\end{figure}

Figure~\ref{fig:cejaf}(a) shows the effect of disorder on the CE phase 
for various strengths $\Delta$ and with $\lambda=0$. The results present interesting similarities
with those obtained for the A-type AF phase in 3D: a conducting system is rapidly stabilized
by increasing $\Delta$ and this phenomenon occurs in a non-monotonic way, with fairly small $\Delta$'s 
having the largest effect in the conductance. At the lowest temperatures the system has an
insulating behavior that becomes metallic at intermediate temperatures and finally it achieves
constant conductance at the high temperature paramagnetic regime.\cite{re:tomioka95}
It is interesting to observe that increasing $\lambda$,
the rapid transformation of the insulator into a metal no longer occurs, as shown in 
Fig.~\ref{fig:cejaf}(b) for the case $\lambda=1$. While a nonzero conductance is reached at low temperatures,
its value is substantially smaller than in the $\lambda$=0 case. This effect is more pronounced
in Fig.~\ref{fig:l8x8cvsd}(a), where the conductance at low temperatures $T$=0.01 is shown. The 
large difference between the small and large $\lambda$ regimes is obvious: at small $\lambda$ the
fragility of the CE state is much more notorious. Also note the similarity of the conductance vs. $\Delta$
plot for the 2D CE state with the results for the 3D A-type AF state. 
We believe the reason is that both
these states
are the closest to the FM metallic state in parameter space, as explained in more detail below.

Finally, the clear correlation between the spin orientations\footnote{Here and in
the following sections, Monte Carlo ``snapshots'' are used to illustrate the prevailing states of the system.
although, of course, the Monte Carlo method can only calculate averages.} at each link
and the value of $J_{\rm AF}$ at that particular link is shown in
Fig.~\ref{fig:l8x8cvsd}(b,c) at large $\Delta$. As in 3D, percolative processes favor a poor metallic
behavior in this regime. But note that 50\% is the critical number of links for 2D percolation and,
as a consequence, the 2D large $\Delta$ regime is only at the verge of metallicity. This explains the
 smaller values in conductance at large $\Delta$ between 2D and 3D.
 The state induced by the disorder acting on the 2D CE state must be similar
 to the ``CE glass'' studied in Ref.~\onlinecite{re:aliaga03} before.

\section{Effect of Disorder on the Magneto-Resistance} \label{sec:magnetoresistance}

The fragility of the insulating states in the $\lambda$=0 regime upon the introduction
of disorder, suggests that other perturbations may also rapidly transform these states
from insulating to metal. In particular, the effect of magnetic fields, $B$, is important
for potential applications and to improve our understanding of the CMR effect of manganites.
To carry out this study, we will proceed sequentially through the phases and parameters 
described before and summarized in Figs.~\ref{fig:phase3dsketch} and \ref{fig:phase2dsketch}. 
The magneto-resistance here will be defined
by $[(R(0)-R(B))/R(0)]\times 100$, where $R(B)$ is the resistance (1/conductance) at the magnetic field $B$.  
The magnetic field is applied in the $z$ direction, same
orientation as the A and CE phases, 
and we have chosen the intensities of $B$=$0.05$ and $B$=$0.10$ for our studies.

\subsection{A-type AF phase}

\begin{figure}
\centering{
\includegraphics[width=8cm,clip]{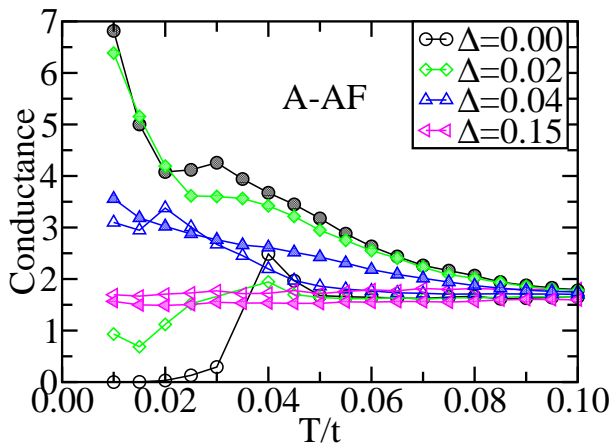}
}
\caption{\label{fig:atypeb} 
Influence of a magnetic field on the transport properties of the 3D A-type AF state, 
as discussed in the text. The effect is particularly dramatic at small $\Delta$, including
the clean limit $\Delta$=0. Shown are the conductances vs. $T$ of the
A-type AF phase, at
different disorder strengths $\Delta$.  Filled symbols (empty symbols)
correspond to a  magnetic field $B$=0.1  ($B$=$0.0$).
The calculations were done on a $4^3$ lattice, at $J_{\rm AF}=0.15$ and 
$\lambda$=$0$, and using 5,000 thermalization and 15,000 measurement Monte Carlo steps 
(calculating the conductance every 10
steps). The starting configuration used was an A-type AF phase. 
The disorder is bimodal, $J_{\rm AF}\pm\Delta$, and shown are averages
over 5 disorder configurations. The magnetic field is applied in the
$z$-direction. 
}
\end{figure}

\begin{figure}
\centering{
\includegraphics[width=8cm,clip]{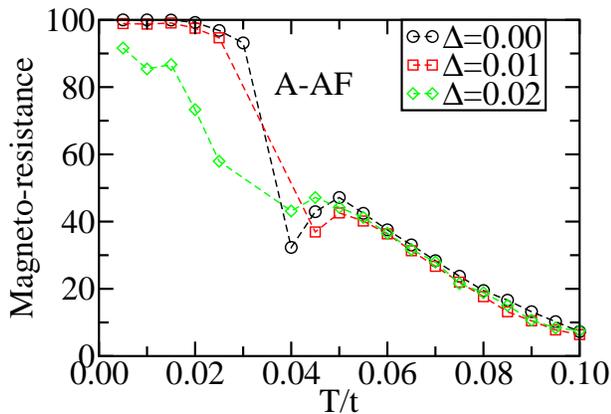}
}
\caption{\label{fig:atypemr} Magneto-resistance (defined in Section~\ref{sec:magnetoresistance})
at $B$=$0.1$ vs. $T$, for the
A-type AF regime in 3D, and considering
several disorder strengths $\Delta$ 
(indicated). The lattice, couplings, Monte Carlo steps, and other conventions
are as in Fig.~\ref{fig:atypeb}. 
}
\end{figure}

Results for the conductance in the 3D A-type AF regime
are given in Fig.~\ref{fig:atypeb}, in the clean and disordered cases, and considering both
$B$=0 and $B$=0.1. It is remarkable that in the clean limit, a relatively small field such
as $B$=0.1 is capable of totally transforming the insulator into a metal. In this case, the
magneto-resistance factor is the largest, very close to 100\%, as shown in Fig.~\ref{fig:atypemr}.
A similar phenomenon occurs at $B$=0.05 (not shown). The magnitude of this effect
remains strong for small $\Delta$s such as 0.02, and then slowly decreases. 
The singular behavior at
temperatures close to $T$=0.03-0.04 is caused by the peaks in conductance at those intermediate
temperatures. 
The focus should be on the low temperature regime, which is the most interesting
in view of the magnitude of the effects. The result at $\Delta$=0 is probably caused
by the close proximity in energy of the FM metallic state, as discussed before in
Ref.\onlinecite{re:aliaga03}. If this is the case, then the transition must be first-order
between the two, and a finite value of $B$ is needed to induce it. From this perspective,
the finite but small $\Delta$ region will be more interesting since even in the limit
of vanishing $B$, it has a substantial magnetic susceptibility, as discussed below. 

\subsection{CE-phase in 3D and 2D}

\begin{figure}
\centering{
\includegraphics[width=8cm,clip]{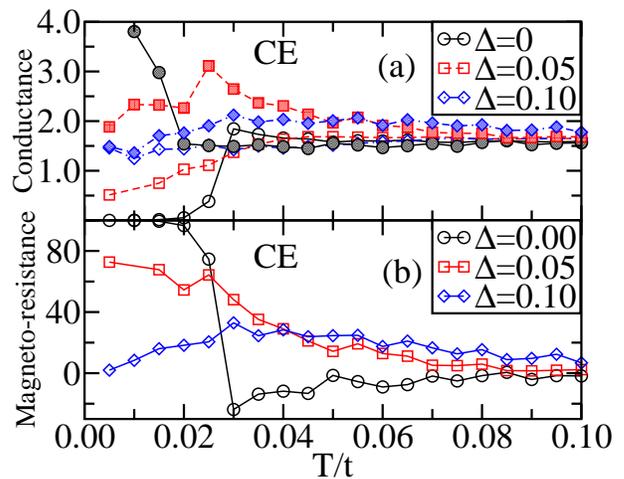}
}
\caption{\label{fig:ce3db} 
(a) Influence of a magnetic field on the transport properties 
in the regime of the CE phase in three dimensions. 
Shown are the conductances vs. $T$ of the CE-type phase,  for
different disorder strengths $\Delta$. Filled symbols (empty symbols)
correspond to  magnetic fields $B$=$0.1$ ($B$=0.0).
Note the large change at low temperatures, and
small and vanishing disorder strength.
Lattice, $\lambda$, Monte Carlo steps, number of quenched disorder configurations
for averages, and other details are as in Fig.\ref{fig:atypeb}. The coupling $J_{\rm AF}$ is 0.2.
(b) Magneto-resistance (defined in Section~\ref{sec:magnetoresistance}) corresponding to the data shown in (a).
}
\end{figure}

Monte Carlo simulations with an applied magnetic field
were also carried out in the regime of the CE phase in three dimensions, and the results are shown
in Fig.~\ref{fig:ce3db} for $B$$=0.1$. As in the previous case of the A-type phase, here also
the CE phase is found to be highly susceptible to 
the application of a magnetic field, and the clean limit insulating state 
quickly becomes metallic. Disordered samples are less
susceptible to the application of a magnetic field, as previously observed for the A-type AF phase.
Both regimes, CE and A-type AF,  present a similar phenomenology.

\begin{figure}
\centering{
\includegraphics[width=8cm,clip]{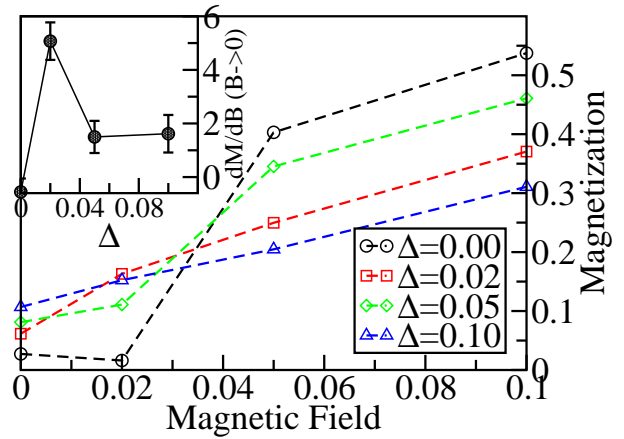}
}
\caption{\label{fig:ce3dmvsb} Magnetization vs.~applied magnetic field, for
various disorder strengths $\Delta$ at $T$=$0.01$. \emph{Inset:}
The magnetic susceptibility $dM/dB|_{B\rightarrow0}$ vs.~disorder strength $\Delta$.
Parameters and other details are as in Fig.~\ref{fig:ce3db}.}
\end{figure}

The analysis of the influence of magnetic fields can be made more precise
by studying the magnetization at small fields $B$. The results are shown
in Fig.~\ref{fig:ce3dmvsb}. At $\Delta$=0, a small
but finite $B$ is needed to render the system ferromagnetic, as previously discussed. 
This is probably
related with the first-order transition reported in previous 
investigations in the clean limit.\cite{re:aliaga03} 
The slope of $M$ vs. $B$ at low fields increases with $\Delta$.
This small $\Delta$ regime is also
the most relevant in some of the cases studied in previous sections in the
absence of magnetic fields.

\begin{figure}
\centering{
\includegraphics[width=7cm,clip]{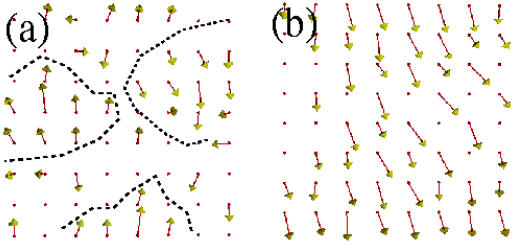}
}
\caption{\label{fig:l8x8snapshot} (a) Monte Carlo ``summed snapshot'' showing the
``clustered spins'' stabilized by a small amount of quenched disorder
at $B$=$0$, for the
case of an 8$\times$8 lattice, $J_{\rm AF}=0.17$, $\lambda$=$0$, and
$\Delta$=$0.02$. In (a), the presence of FM clusters is clear to the eye:
quenched disorder with a small value of $\Delta$ transforms
the original CE state into a state with local ferromagnetism (but not
global). The conductance of this state is zero.
(b) Same as in (a) but for $B$=$0.02$. 
A very small magnetic field is sufficient to
align the pre-formed FM clusters, and simultaneously 
increasing the conductance (to a value $C$=$3$, in this case).
}
\end{figure}

To further investigate the origin of these effects, Monte Carlo
thermalized spin configurations have been analyzed. From each of these configurations
we constructed the ``summed snapshot'', $\sum_{j\in \mathcal{V}_i} \vec{S}_j$, where $\mathcal{V}_i$
represents the vicinity (4 sites in 2D or 6 sites in 3D) of site $i$.
The advantage of this quantity is that it helps identify regions that are FM. If the
spins near $i$ are pointing randomly then the ``summed'' spin tends to cancel.  
In Fig.~\ref{fig:l8x8snapshot}(a), an example is shown using an $8\times8$ lattice
(very similar results were obtained on a $20\times20$ cluster using the TPEM
method described later in this paper). In the absence
of magnetic fields, but including disorder with a small $\Delta$, 
the stabilized spin configuration is
very different from the CE phase with parallel zigzag chains
that is present at $\Delta$=0. With this disorder, ferromagnetic clusters containing many spins with
the same orientation are present. Their random orientation
character gives a nearly zero global magnetization. However, clearly a system
with preformed FM islands must have a high susceptibility to external
magnetic fields, as already argued to be the reason for the manganite CMR effect in previous
studies.\cite{re:dagotto01} When magnetic fields are applied, the FM clusters rapidly orient
their moments creating a global FM state, 
as shown in Fig.~\ref{fig:l8x8snapshot}(b).

\begin{figure}
\centering{
\includegraphics[width=8cm,clip]{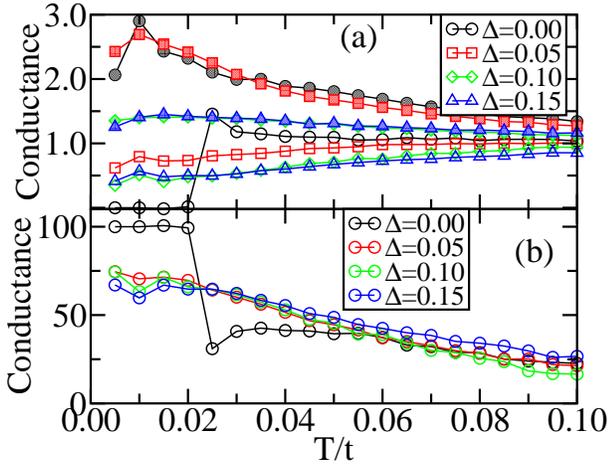}
}
\caption{\label{fig:ce2dlambda0bb} (a) Conductance vs. $T$ of the
two-dimensional CE phase  for
different disorder strengths $\Delta$. Full (empty) symbols correspond to a magnetic field
$B$=$0.1$ ($B$=0). 
The calculation was done on an $8\times8$ lattice at $J_{\rm AF}=0.17$, $\lambda$=$0$, 
and using 5,000 thermalization and 15,000 measurement Monte Carlo steps, and the CE phase as initial
configuration . The disorder is bimodal, $J_{\rm AF}\pm\Delta$, and the average shown is 
over 5 quenched disorder configurations.
(b) Magneto-resistance at $B$=$0.1$ vs. $T$ for the 2D
CE phase, at  different disorder strengths $\Delta$, as indicated. 
The calculation was done using the same lattice, couplings, and other details as in (a).
}
\end{figure}

For completeness, 
the effect of a magnetic field on the CE phase in two dimensions
was also studied. Results for the conductance 
and the magnetoresistance are
in Fig.~\ref{fig:ce2dlambda0bb} for $\lambda$=$0$. The trends are very similar
to those found in three dimensions.

\section{Results using on-site disorder}

In this section the influence of an on-site
disorder (bimodal, of strength $\Delta_{\rm P}$) will be investigated. The term added to the
Hamiltonian is an on-site random energy, $\sum_{i,\gamma} \epsilon_i c^\dagger_{i,\gamma} c_{i,\gamma}$
with $\epsilon_i =\pm \Delta_{\rm P}$, to be contrasted with the link randomness
considered in the previous sections. It will be shown that the conclusions regarding the fragility
of the insulating states at $\lambda$=0 remain valid, 
even if the source of disorder is changed.

\begin{figure}
\centering{
\includegraphics[width=8cm,clip]{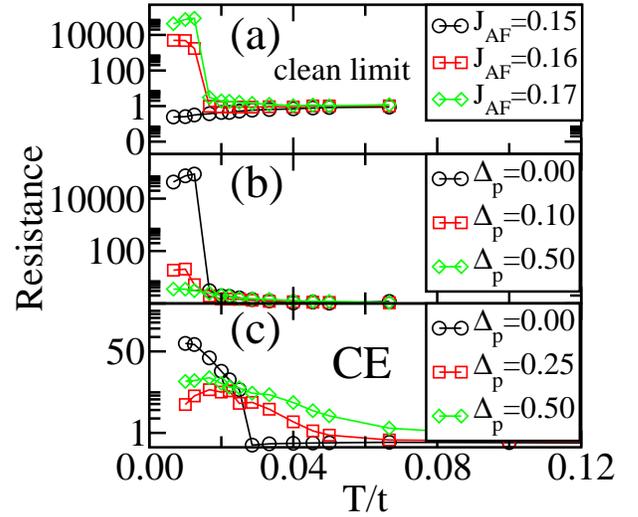}
}
\caption{(a) Resistance vs. temperature for a 12$\times$12 lattice in the 
clean limit $\Delta$=0, $\lambda=0$, and considering 
$J_{\rm AF}=0.15$ (squares), $J_{\rm AF}=0.16$ (circles), and $J_{\rm AF}=0.17$ (triangles). 
The figure illustrates the metallic character of the $J_{\rm AF}=0.15$ state (with a FM
ground state),  and the insulating properties of the $J_{\rm AF}=0.16$ and 0.17 samples (with a
CE ground state).
The Monte Carlo simulation was carried out starting with a random state at each temperature, 
using a warming up of 10,000 steps, followed by an additional 10,000 steps to measure properties. 
(b) Same as (a) at $J_{\rm AF}=0.17$, 
but now considering the on-site disorder strengths $\Delta_{\rm P}$ indicated, and averaging
over 3 different disorder configurations.
The resistance of the clean sample is reduced several orders of magnitude by the addition
of the on-site potential disorder, similarly as it occurs for link disorder.
(c) Same as (a), but on a $4^3$ lattice and with $J_{\rm AF}=0.20$.
\label{fig:hifigfake1a}}
\end{figure}

Figure~\ref{fig:hifigfake1a}(a) shows the resistance obtained in the clean limit,
parametric with $J_{\rm AF}$, using a 2D cluster. This result is the analog of the conductance
plots in the clean limit presented in previous sections. Varying $J_{\rm AF}$,
an abrupt transition from a metal to an insulator is obtained at low temperatures, as expected.
Let us now introduce on-site disorder. The results are in Fig.~\ref{fig:hifigfake1a}(b). It is clear
that the insulating state dramatically reduces its resistance with increasing $\Delta_{\rm P}$, 
transforming into a bad metallic state, as found in the case of link disorder. The changes in
the resistance are of several orders of magnitude. The same occurs 
in three dimensions, Fig.~\ref{fig:hifigfake1a}(c), where in addition it is shown that the 
resistance has a minimum at intermediate values of the disorder strength $\Delta_{\rm P}$,
as it occurs for link disorder. Although the substantial numerical effort involved in the simulations
presented in this manuscript prevents us from discussing the site disorder case in as much detailed
as the case of link disorder, nevertheless it is safe to conclude that site disorder 
has a phenomenology similar to that of link disorder, in the sense that by increasing $\Delta_{\rm P}$
a conductor is obtained.

\section{Finite-Size Effects and TPEM Technique}

\begin{figure}
\centering{
\includegraphics[width=8cm,clip]{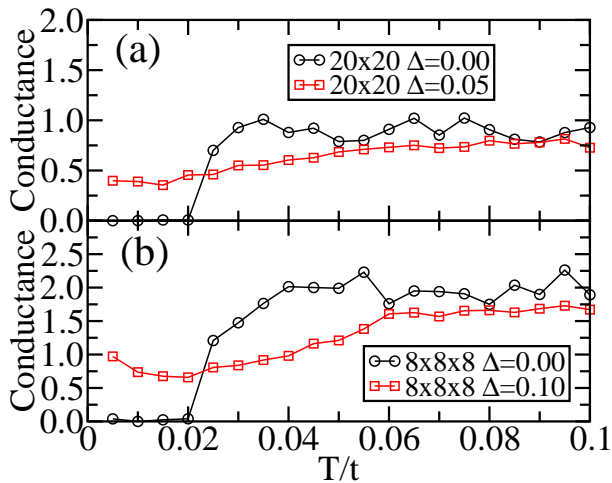}
}
\caption{\label{fig:tpem} 
(a) Effect of quenched disorder on the two-dimensional CE phase, running the simulation with the TPEM.
Conductance vs. $T$ of the CE phase on a 20$\times$20 lattice
calculated with the TPEM method using $M$=$30$ moments, and the TPEM parameters $\epsilon_{\rm pr}$=$10^{-6}$
and $\epsilon_{\rm tr}$=$10^{-6}$. The coupling 
was $J_{\rm AF}=0.17$, density $n$=$0.5$,
and 2,000 for thermalization and 1,000 for measurement Monte Carlo steps were carried out. 
The CE phase shows zero
conductance at low temperature in the clean limit. This transforms into a disordered metal with
finite conductance when link quenched disorder of strength $\Delta$=$0.05$ is added.
(b)  Effect of disorder on the three-dimensional CE phase, studied with TPEM.
Details are the same as in (a) but on an $8^3$ lattice and at $J_{\rm AF}=0.20$.
}
\end{figure}

The investigations reported in the previous sections 
have been carried out using intermediate size clusters, at the verge of the sizes that can be
handled with the exact diagonalization treatment of the fermionic sector. Fortunately, the systematic
effects observed, and similarities of results between different quenched disorder configurations,
dimensionality, and insulating phases strongly point toward a robust effect, and we are confident
that the essence of the phenomenon has been unveiled. Nevertheless, only the
study of larger clusters would convincingly rule out spurious size effects in our investigations.
Fortunately, recent algorithm developments have led to the use of the ``Truncated Polynomial
Expansion Method'' (TPEM)\cite{re:motome99,re:furukawa01,re:furukawa03,re:motome03b} 
as an alternative to conventional Monte Carlo methods where the fermionic
sector is treated exactly. The advantage of the TPEM is that its CPU time grows linearly with
the number of sites $N$, while the canonical approach used in the previous sections demands an effort
that scales as $N^4$. This advantage is not without penalization: the coefficient in front of
$N$ in the TPEM scaling is very large. For this reason, we studied few large lattices,
 mainly to check if the results on small systems are preserved at larger $N$ values.
After confirming that the TPEM reproduces the results of small systems accurately (not shown
but see Ref.~\onlinecite{re:sen05}),
studies were carried out using $20\times20$ and $8\times8\times8$ clusters.
In Fig.~\ref{fig:tpem}, the results are shown, both
in the clean limit and introducing disorder. In both cases, at low temperatures the conductance
changes from exactly zero to a finite number, as observed in previous sections. While this is not
a final proof, at least it has been confirmed that increasing the lattice sizes does not lead
to dramatic changes in our conclusions, and the insulator to metal transition with increasing $\Delta$
seems to be a solid property of the CE states of manganites at small electron-phonon coupling.

\section{Discussion and Conclusions}

In this paper, the influence of small amounts of quenched disorder on the transport properties
of manganite models was investigated. The focus has been on the insulating phases that are
stabilized without electron-phonon coupling,\cite{re:hotta00} namely those that have insulating properties
due to geometrical considerations and frustrating competing tendencies between FM and AF
states.

\begin{figure}
\centering{
\includegraphics[width=8.5cm,clip]{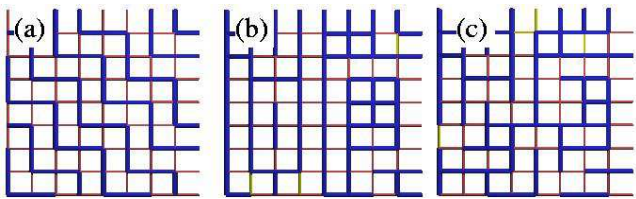}
}
\caption{\label{fig:snapshots}
Results of Monte Carlo simulations that summarize the main conclusions
of this manuscript (see text for discussion).
Shown are the signs of the classical spin correlation 
$\vec{S}_i\cdot \vec{S}_j$ 
considering nearest-neighbor sites $i,j$,
indicated as red/thin (negative; AF) and blue/thick (positive; FM) connections. The
lattice is 8$\times$8, $J_{\rm AF}=0.17$, and the disorder strength is (a)
$\Delta$=$0.01$, (b) $\Delta$=$0.02$, (c) $\Delta$=$0.08$. 
In (a), the disorder is
too weak to change the CE spin pattern, which is clear in the result. 
However, in (b) the CE phase is replaced by
long ferromagnetic (metallic) chains, which in this case are vertical.
These chains provide conductance channels in the vertical direction.
An equal number of horizontally conducting systems are found when analyzing
many quenched disorder realizations, thus restoring rotational invariance after
averaging.
In (c), the state is
more reminiscent of the high disorder regime,
with shorter chains. At large $\Delta$, the spins follow the disorder as
explained already in Section~\ref{subsec:ce2d}. 
Of the three configurations, the largest conductance is obtained in (b).}
\end{figure}

The study reported here has identified regimes where the insulating phases, the CE and A-type AF states,
are very susceptible to the introduction of disorder. For instance, the
A-type AF phase in 3D and the CE-type phase in 2D, they both present a
conductance that sharply increases with increasing $\Delta$ from zero,
and then it settles to a smaller value at large $\Delta$. The reason appears
to be explained with the type of spin arrangements 
that become stabilized in the presence of quenched disorder. Typical results are shown in
Fig.~\ref{fig:snapshots}, for the case of the 2D CE state. 
In (a), the CE pattern is nicely reproduced by the
Monte Carlo simulation at zero and very small $\Delta$. However, in (b), a key observation
is that $lines$ with classical
spins align ferromagnetically are stabilized along the vertical direction for a small $\Delta$. 
For instance, the first and fourth column from the left are all FM. 
The second, fourth and fifth
columns are almost entirely FM, with the exception of only one link. More
quantitatively, there are 50 links FM in the vertical direction but only 27
FM links in the horizontal direction (ratio approximately 2). These lines provide conducting paths
for the charge, rendering the system conducting in the direction of those lines.
Even though the conductance is not always measured along these lines, the average over all
directions is still large. 
This is to be contrasted with
the regime shown in (c)
for large values of $\Delta$, where the ratio between FM links along
the vertical and horizontal axes is approximately 1, as is also in the clean limit. At least
for 3D, this type of configurations is also conducting, as explained with a percolative picture
in previous sections.
These tendencies were observed in numerous realizations of disorder, with the only
difference between them being that the dominance of the horizontal or vertical
lines has equal chance. Although most of the investigations reported here 
were obtained using intermediate size clusters,
some larger systems were studied with the TPEM and the results remain similar. Nevertheless,
a more sophisticated analysis is, of course, always desirable to confirm our conjectures.
Hopefully, our effort will induce further work in this challenging area of research.

The origin of the FM lines that are rapidly stabilized by disorder must
be the close proximity in energy of the full FM state. While the CE state
prefers zigzag chains (which are band insulators\cite{re:hotta00}) to minimize the energy, 
the FM metallic state prefers straight
lines since they are metallic. So this striped state can be considered
as a compromise between the CE zigzag insulating and the FM uniform metallic states of the clean limit.
Note that it is also possible to imagine a ``clustered'' state as a compromise
of these two tendencies, and indeed in some circumstances this type of states was
observed in our simulations, as it was shown in Fig.~\ref{fig:l8x8snapshot}(a). But this state
has zero conductance, while the striped state of Fig.~\ref{fig:snapshots}(b) has a finite
conductance and, thus, it is responsible for the rapid insulator to metal transition with
increasing $\Delta$.

We believe the conclusions reached in this paper have consequences beyond the case of Mn oxides.
Clearly, the entire physics discussed arises from competition of phases, which is a common phenomenon
in transition metal oxides. This phase competition is already
present in the clean limit, since the CE and A-type AF phases appear in parameter space in narrow
regions of the phase diagram, as a compromise between the FM and AF dominant tendencies. The same
was recently shown to occur in between a $d$-wave superconductor and an antiferromagnet\cite{re:alvarez05c}
in a study of a phenomenological model for cuprates: in this case a striped phase was found as a compromise
of the two tendencies in some regions of parameter space.
It should not be surprising that the intermediate phases stabilized by phase competition 
are fragile, namely they change dramatically
when influenced by disorder or external fields. This fragility manifests in the fairly exotic states
that are stabilized when small quenched disorder is introduced: for the case of the manganites, 
these are states with either FM clusters or
FM lines as reported here, having in common regions where ferromagnetism and metallicity dominates locally.
That this occurs at values of $\Delta$ as small as 0.02 is the most surprising result. This high
susceptibility to disorder should exist in many other complex oxides as well.
When these giant effects are compared with other insulator to metal transitions induced by disorder
(see Section \ref{sec:intro})
it appears that the cases discussed here have a unique character, not previously 
discussed in the literature. 
The consequences of these results for our understanding of manganites and the CMR effect
seem clear: it is the fragile nature of the insulator to the introduction of 
external perturbations (quenched disorder, magnetic fields, electric fields, \ldots) 
that may cause the well-known giant responses of Mn oxides.

\begin{acknowledgments}
Conversations with V.~Dobrosavljevic are gratefully acknowledged.
G.~A. is supported by the Eugene P. Wigner Fellowship Program at Oak Ridge National Laboratory (ORNL). 
H.~A., C.~S. and E.~D. are supported by the NSF grant
DMR-0443144. H.~A. is also supported by the LDRD program at ORNL.
We acknowledge the National Center for Computational Sciences for 
providing most of the computational resources. 
ORNL is managed by UT-Battelle, LLC, for the U.S. Department of 
Energy under Contract DE-AC05-00OR22725. 
\end{acknowledgments}

\bibliography{thesis}

\end{document}